\documentclass[aps,prl,twocolumn,showpacs,amsmath,amssymb,floatfix,superscriptaddress]{revtex4}
\usepackage{graphicx}
\usepackage{dcolumn}
\usepackage{bm}
\usepackage{graphicx}
\usepackage{caption}
\usepackage{color}
\usepackage{amsmath, amsthm, amssymb}
\usepackage[modulo]{lineno}
\usepackage[usenames,dvipsnames]{xcolor}
\usepackage{cancel}
\newcommand{\ba}{\begin{eqnarray}}
\newcommand{\ea}{\end{eqnarray}}

\begin{document}
\title{Measurement of the electronic thermal conductance channels and heat capacity of graphene at low temperature}
\author{Kin Chung Fong}
\author{Emma Wollman}
\author{Harish Ravi}
\affiliation{Kavli Nanoscience Institute, California Institute of Technology, MC 128-95, Pasadena, California 91125}

\author{Wei Chen}
\author{Aashish A. Clerk}
\affiliation{Department of Physics, McGill University, Montreal, Canada H3A 2T8}

\author{M.D. Shaw}
\author{H.G. Leduc}
\affiliation{Jet Propulsion Laboratory, California Institute of Technology, Pasadena, California 91109, USA}

\author{K.C. Schwab}
\affiliation{Kavli Nanoscience Institute, California Institute of Technology, MC 128-95, Pasadena, California 91125}

\date{\today}


\begin{abstract}
{The ability to transport energy is a fundamental property of the two-dimensional Dirac fermions in graphene. Electronic thermal transport in this system is relatively unexplored and is expected to show unique fundamental properties and to play an important role in future applications of graphene, including opto-electronics, plasmonics, and ultra-sensitive bolometry. Here we present measurements of bipolar, electron-diffusion and electron-phonon thermal conductances, and infer the electronic specific heat, with a minimum value of 10 $k_{\rm{B}}$ ($10^{-22}$ JK$^{-1}$) per square micron. We test the validity of the Wiedemann-Franz law and find the Lorenz number equals $1.32\times(\pi^2/3)(k_{\rm{B}}/e)^2$.  The electron-phonon thermal conductance has a temperature power law $T^2$ at high doping levels, and the coupling parameter is consistent with recent theory, indicating its enhancement by impurity scattering.  We demonstrate control of the thermal conductance by electrical gating and by suppressing the diffusion channel using superconducting electrodes, which sets the stage for future graphene-based single microwave photon detection.\\}
\end{abstract}

\maketitle

\section{Introduction}
Electrical transport in graphene has attracted much attention due to the pseudo-chiral and relativistic nature of the band structure\cite{Zhang:2005p1950, Novoselov:2005p1955}. Since both electrons and holes carry energy as well as charge, the thermal transport of Dirac Fermions in two dimensions is expected to be as fascinating as its electrical counterpart. Theorists have suggested a number of intriguing possibilities: the relativistic hydrodynamics of a Coulomb-interacting electron-hole plasma may result in deviations from the Fermi-liquid values of the Mott relation,  Wiedemann-Franz ratio\cite{Muller:2008p2222, Foster:2009p2216},  and electronic specific heat\cite{Vafek:2007p2071} . Thermal transport measurements may reveal the physics of a neutral mode in the fractional quantum Hall effect\cite{Bid:2013p2412}. The thermal properties of the electron gas are also critical to graphene-based device applications\cite{Balandin:2011p2120, Pop:2012p2334}, as they impact photodetector performance\cite{Gabor:2011p2063}, place fundamental limits on the mobility of charge carriers\cite{Morozov:2008p2056}, and set the sensitivity of terahertz and microwave-frequency bolometers\cite{Fong:2012p2256, Vora:12p2211, Yan:2012p2292}, which promise single-photon resolution due to the expected minute specific heat\cite{Fong:2012p2256, McKitterick:2013p2396}.

We present measurements of the bipolar thermal conductance over a temperature range of 300 mK to 100 K, using three different sample configurations (described below).  For temperatures below $\sim$1 K, we identify the thermal transport due to electron diffusion, $G_{\rm{wf}}$, test the Wiedemann-Franz (wf) law, and infer the electronic heat capacity, with a minimum value of 10$^{-20}$ J/K at 300 mK, which is 9 times smaller than the previous record\cite{Wei:2008p2106}. For higher temperatures, we  measure the thermal conductance due to phonon emission, $G_{\rm{ep}}$, while varying the charge density.  There has been recent theory\cite{Chen:2012p2291} which explores the effects of electronic disorder on the electron-phonon (ep) coupling mechanism and predicts a substantial modification in comparison to earlier theory in the clean limit\cite{Bistritzer:2009p1891, Tse:2009p1892, Kubakaddi:2009p1895, Viljas:2010p1976}; the disordered limit is defined by $\lambda_{\rm{p}}\gg l_{\rm{e}}$, where $\lambda_{\rm{p}}=hs/(k_{\rm{B}}T)$ is the dominant thermal phonon wavelength, $l_{\rm{e}}$ is the electron mean free path, $s$ is the sound velocity of graphene acoustic phonons, and $k_{\rm{B}}$ and $h$ are Boltzmann's and Planck's constant, respectively. We present measurements which both confirm the effect of the disorder and the nature of the ep coupling (scalar, vector, screened or unscreened).

Previous thermal studies of graphene have been limited to measurements of thermoelectric power\cite{Zuev:2009p1831, Wei:2009p2105, Checkelsky:2009p2213}, or to measurements of thermal conductance taken at temperatures above the Bloch-Gr\"{u}neisen temperature\cite{Betz:2012p2336, Graham:2012p2258}, at the charge neutrality point (CNP)\cite{Fong:2012p2256}, or without considering the effects of disorder\cite{Betz:2013p2257}. Significant discrepancies between the theoretical\cite{Bistritzer:2009p1891, Tse:2009p1892, Kubakaddi:2009p1895, Viljas:2010p1976} and measured values\cite{Betz:2013p2257} of both the ep-coupling temperature power law and the coupling constant are found in some of these experiments. 

\section{Experiments}
We probe thermal transport of the electrons in graphene by applying Joule heating and measuring the electron temperature utilizing ultra-sensitive, microwave frequency Johnson noise thermometry with a sensitivity of 2 mK/$\sqrt{\rm{Hz}}$\cite{Spietz:2003p2080, Fong:2012p2256} [see Fig. 1a and Supplementary Material (SM)].  Fig. 1b shows the expected thermal model of the electron gas. With a typical coupling bandwidth of 80 MHz to the graphene\cite{Fong:2012p2256}, one-dimensional thermal transport\cite{Schwab:2000p974} through black body radiation, $G_{\rm{rad}}$ ($\simeq$10$^{-15}$ pW/K) $\ll (G_{\rm{wf}}$, $G_{\rm{ep}})$ is expected to be negligible in this experiment.  We assume both the electrodes and lattice are in thermal equilibrium with the sample stage as the ep coupling in normal metal\cite{Roukes:1985p1900} and the boundary thermal conductance of the SiO$_2$-graphene interface\cite{Chen:2009p2116} are large compared to the $G_{\rm{wf}}$ and $G_{\rm{ep}}$ thermal channels. Three devices with different electrodes and gating materials (see Tab.~1 and SI) are measured in two cryostats to cover the entire sample temperature range: 0.3-1.5 K and $T>$ 1.5 K. For all three samples, the device length is much longer than the inelastic scattering length, $l_e$, which avoids any issues of electron shot noise\cite{Steinbach:1996p1728, Danneau:2008p196802, DiCarlo:2008p156801}. For charge densities which can be reached with our experiment, $n=10^{11} -10^{13}$ cm$^{-2}$, the transition from ep to electron-diffusion cooling is expected to occur at $\sim$1 K and should be apparent due to the difference in temperature dependence of the thermal conductance:  $G_{\rm{ep}}$ and $G_{\rm{wf}}$ are expected to depend on temperature as $T^{\delta-1}$ (with $\delta \geq 3$ typically) and $T$, respectively.  

With Joule heating, $\dot Q$, applied to the electron gas, the electron temperature, $T_e$, is expected to follow the two-dimensional heat transfer differential equation: \ba \dot q = -\mathbf{\nabla}\cdot\left(\kappa_{\rm{wf}}\mathbf{\nabla}T_{\rm{e}}\right) + \Sigma_{\rm{ep}}\left(T_{\rm{e}}^\delta- T_{\rm{p}}^\delta\right) \label{eqn:ThermalDiffEqn}\ea where $\dot q= E^2 /\rho$ is the local Joule heating (such that $\int \dot q d^2\mathbf{r} = \dot Q$), E is the local electric field, $\rho$ is the electrical resistivity, $\kappa_{\rm{wf}}$ is the thermal conductivity due to electronic diffusion, $\Sigma_{\rm{ep}}$ is the ep coupling parameter, and $T_{\rm{p}}$ is the local phonon temperature. On the right hand side, the first term describes diffusive cooling through the electron gas, while the second term describes cooling by phonon emission. In a Fermi liquid, the local Joule heating and diffusive cooling are connected through the wf law,  $\kappa_{\rm{wf}} = \mathcal{L}_0T_e/\rho$, where $\mathcal{L}_0$ is the Lorenz number given by $(\pi^2/3)(k_{\rm{B}}/e)^2$ .   Since the temperature of the sample will not be uniform (the middle will have a higher temperature than the leads) and we measure the average electron noise temperature, the wf relationship will be modified to $G_{\rm{wf}} = \alpha\mathcal{L}_0T_{\rm{p}}/R$, where $R$ is the graphene resistance and $\alpha=12$ (see SM for discussion.)\cite{Prober:1993p2102}

By computing the ratio of $\dot Q$ to the measured increase in average electron temperature with $\Delta T_{\rm{e}} / T_{\rm{p}}\ll 1$,  we determine the thermal conductance, $G_{\rm{th}}$. Fig.~2a shows the results from device D1 (gold leads, top-gated) at various charge carrier densities. There is a clear transition from a quadratic to linear temperature dependence at $\sim$ 1K, which is expected and can be understood as $G_{\rm{wf}}$ dominating at low temperatures (Fig.~2b).  We test the Wiedemann-Franz law for two-dimensional Dirac Fermions by plotting $G_{\rm{wf}}$ versus $T/R$ (Fig.~2c) such that the slope is $\alpha \mathcal{L}_0$. We also note that this $G_{\rm{wf}}$ is not equal to zero at $T/R$ = 0, which at this point is not understood. Fig.~2d shows the measured Lorenz number at different densities. The averaged Lorenz number values for electron and hole doping are $1.34\mathcal{L}_0$  and $1.26\mathcal{L}_0$ , respectively.

Our measured Lorenz number is 35\% higher than the Fermi liquid value, 17\% higher than the measured value in graphite\cite{Holland:1966p903}, and comparable to values obtained in other materials\cite{Kumar:1993p2263}. While the electron-electron (e-e) interaction may modify the Lorenz number in a material\cite{Kumar:1993p2263, Muller:2008p2222, Foster:2009p2216}, other effects such as contact resistance and contributions due to graphene under the contacts could contribute to errors in our calculation of the Lorenz number. Four-point probe measurements of the thermal conductance may solve this problem in the future. We also observe an increase of the Lorenz number near the CNP (Fig.~2d). At the Dirac point and at high temperatures where $E_{\rm{F}}\ll k_{\rm{B}}T$ (where $E_{\rm{F}}$ is the Fermi energy), theory predicts that the system becomes quantum critical and the interaction between massless electrons and holes enhances the Lorenz number\cite{Muller:2008p2222, Foster:2009p2216}. The same theory also expects that a deviation from the Fermi liquid value is possible in the case where $E_{\rm{F}}\gg k_{\rm{B}}T$, as is true for our impurity-limited samples, but only if the screening is weak\cite{Muller:2008p2222}. Further experiments in cleaner samples are needed to understand this anomalous behavior in the Lorenz number as well as the offset in Fig.~2c. 

The electronic specific heat capacity, $c_{\rm{e}}$, in graphene can be determined by applying a two-dimensional kinetic model: $\kappa_{\rm{wf}} = (1/2)c_{\rm{e}}v_{\rm{F}}l_{\rm{e}}$, where $C_{\rm{e}}=Ac_e$ is the total electronic heat capacity, $v_{\rm{F}}$ is the Fermi velocity, and $A$ is the device area. Using $l_{\rm{e}} = 32$ nm (Fig.~1e), we plot $C_{\rm{e}}$ on the right hand side of the y-axis in Fig.~2b. Since $G_{\rm{wf}}\propto T$ and $l_{\rm{e}}$ does not depend on temperature significantly due to impurity scattering, the measured specific heat is linear in $T$ for all densities. This agrees with theories for $|E_{\rm{F}}|\gg k_{\rm{B}}T$, as the minimum $|E_{\rm{F}}|$ of our samples is limited by impurity doping. For $|E_{\rm{F}}|\ll k_{\rm{B}}T$, which is not accessible in this experiment, the specific heat is expected to be proportional to $T^2$ in the case of massless Dirac Fermions\cite{Wang:2012p2431}. The smallest specific heat attained near the CNP is merely 10 $k_{\rm{B}}~\mu$m$^{-2}$ or 1000 $k_{\rm{B}}$ for the whole sample, a factor of 9 smaller than the inferred value in state-of-the-art nano-wires used for bolometry\cite{Wei:2008p2106}. This value is also consistent with our earlier result at 5 K estimated using a bolometric mixing effect\cite{Fong:2012p2256}.

At higher sample temperatures or for large Joule heating, the thermal conductance changes its temperature power law behavior (Fig.~2b) as the dominant cooling mechanism switches from $G_{\rm{wf}}$ to $G_{\rm{ep}}$; the cross-over temperature is given by $(\alpha\mathcal{L}_0/\delta RA\Sigma_{\rm{ep}})^{1/(\delta-2)}$. In this regime, Eq.~\ref{eqn:ThermalDiffEqn} reduces to\cite{Kubakaddi:2009p1895, Viljas:2010p1976, Chen:2012p2291}: \ba \dot Q = A\Sigma_{\rm{ep}}(T_{\rm{e}}^\delta-T_{\rm{p}}^\delta)\text{.}\label{eqn:HeatTransfer1}\ea Using a dc current bias, Fig.~3a plots the measured $T_{\rm{e}}$ versus the applied Joule heating power. For large heating powers, the electron temperature converges to $(\dot{Q}/A\Sigma_{\rm{ep}})^{1/\delta}$, independent of the initial temperature. The solid lines in Fig.~3a are the best fit to $T_{\rm{e}} = (\dot{Q}/A\Sigma_{\rm{ep}}+T_{\rm{p}}^3)^{1/3}$, establishing  $\delta=3$.   In light of recent theory\cite{Chen:2012p2291}, our experimental data suggest that the ep heat transfer in the disordered limit is primarily due to a weakly-screened deformation potential, consistent with recent electrical transport measurements\cite{Efetov:2010p2065}. The simplified physical scenario is that impurity scattering in disordered graphene prolongs the ep interaction time and thus enhances the emission rates. Recent investigations\cite{Song:2012p2289, Betz:2012p2336, Graham:2012p2258} suggest that the same power law, $\delta =3$, may also govern the disorder-assisted cooling rate of hot electrons for $T > T_{\rm{BG}}$, but through a different mechanism.

For the three devices we fabricated and measured, all show $\delta \simeq 3$ except for device D2 (gold leads, back-gated) at temperatures above 50 K and device D1 near the CNP (see Fig.~4b). In both circumstances, the temperature power law increases from $\delta = 3$ to 4. This behavior of D1 near the CNP, where transport is expected to be dominated by disorder and charge puddles, is surprisingly consistent with theoretical expectations\cite{Bistritzer:2009p1891, Tse:2009p1892, Kubakaddi:2009p1895, Viljas:2010p1976} in the clean limit, as reported in Ref 11. In this device, as the charge carrier density decreases to 10$^{11}$ cm$^{-2}$ near the CNP, the screening length grows to 50 nm which is comparable to the distance to the nearby metallic top gate (100 nm, dielectric constant of 4). We speculate this screening of the metallic gate may impact the ep coupling. Moreover, near the CNP, the impurity scattering is long-range in nature and $k_{\rm{F}}l_{\rm{e}} < 1$, which is outside the regime of validity of the existing ep coupling theory for disordered graphene\cite{Chen:2012p2291}. More experiments and theory are required to understand the nature of ep coupling at low charge carrier density. This is of particular importance to graphene-based bolometry as the ultimate sensitivity is expected to be limited by the ep coupling at the lowest carrier densities\cite{Fong:2012p2256, McKitterick:2013p2396}.

We can further explore Eq.~\ref{eqn:HeatTransfer1} by measuring the differential thermal conductance at different temperatures and carrier densities using a small ac current bias. If $T_{\rm{e}}-T_{\rm{p}}\ll T_{\rm{p}}$, the ep thermal conductance is $\delta A\Sigma_{\rm{ep}}T_{\rm{p}}^{\delta-1}$. Fig.~3b shows $G_{\rm{ep}}$ as function of carrier density for device D3 (superconducting leads, back-gated). Fig.~2b and 3c, for devices D1 and D3 respectively, show that $G_{\rm{th}}$ is limited by the ep thermal conductance at T $>$ 1.5 K with a power law $\delta\simeq 3$ for both devices. However, since the electrons in a superconductor have negligible entropy and do not conduct thermally, $G_{\rm{th}}$ of device D3 with NbTiN electrodes is not limited by $G_{\rm{wf}}$ at low temperatures. The dashed line in Fig.~3c is the calculated $G_{\rm{wf}}$, similar to the dashed line in Fig.~2b. NbTiN is used because of its higher transition temperature (14 K) to avoid e-e interaction that may promote hot electrons over the superconducting bandgap\cite{Voutilainen:2011p2043}. The ep data at 0.4 K demonstrate the suppression of heat diffusion by roughly 80\%. 

We can obtain $\Sigma_{\rm{ep}}$ and $\delta$ by fitting $G_{\rm{ep}}$ as function of lattice temperature at a constant carrier density. Results are plotted in Fig.~4. For devices D2 and D3, we find $\delta$ values of approximately 3.0 and 2.8, respectively. Near the CNP, $\Sigma_{\rm{ep}}$ has a minimum, but is not vanishing. Furthermore, the fitted $\Sigma_{\rm{ep}}$ values vary by an order of magnitude across all three devices. This variation and also the discrete jump in measured $G_{\rm{th}}$ after thermal cycling in Fig.~3c indicate that the underlying ep coupling mechanism is strongly modified by disorder.

We compare the theory of ep coupling in disordered graphene to both the measured electrical and ep thermal transport data using\cite{Chen:2012p2291}:\ba\Sigma_{\rm{ep}} = \frac{2\zeta(3)}{\pi^2}\frac{E_{\rm{F}}}{v_{\rm{F}}^3\rho_M}\frac{\mathcal{D}^2k_{\rm{B}}^3}{\hbar^4l_{\rm{e}}s^2}\ea where $\mathcal{D}$ is the deformation potential and $\rho_M$ is the mass density of the graphene sheet. This theory assumes a short-range scattering impurity potential and $k_{\rm{F}} l_{\rm{e}} > 1$. The estimated values of $\mathcal{D}$ for devices D1, D2, and D3, at a charge density of $n\simeq 3.5\times 10^{12}$ cm$^{-2}$, are 19, 23, and 51 eV respectively (Tab.~1). The considerable scatter in the inferred values of $\mathcal{D}$ is consistent with the range of values obtained from electrical transport measurements\cite{Bolotin:2008p1770, Chen:2008p2054, Dean:2010p2055, Efetov:2010p2065}. Fig.~4b inset shows the inferred deformation potential values in both thermal and electrical experiments reported in the literature\cite{Fong:2012p2256, Bolotin:2008p1770, Dean:2010p2055, Efetov:2010p2065, Chen:2008p2054, Betz:2012p2336, Betz:2013p2257, Graham:2012p2258}. The wide scatter suggests that some factors may not be captured in the ep coupling theory for disordered graphene, such as long-range impurity scattering, substrate-induced effects, or surface-acoustic phonons\cite{Kaasbjerg:2012p2343}.

\section{Conclusions}
In this report, we investigate the bipolar thermal conductance of graphene in both the electron diffusion and electron-phonon regimes.  We find that the ep coupling is strongly modified by electronic disorder and is consistent with scalar coupling in the weak screening limit.\cite{Chen:2012p2291}  ep coupling in the disordered limit is especially relevant for ultra-sensitive device applications at low temperatures, since even the cleanest samples\cite{Bolotin:2008p1770, Dean:2010p2055} yet reported ($l_e > 1~\mu$m) would cross into the disordered limit for temperatures below 1 K.   This experiment has validated the wf law for two-dimensional Dirac fermions. It may be possible to study many-body physics in this system through more precise measurements of the Lorenz ratio and with cleaner samples near the CNP. The electronic specific heat inferred through the electron diffusion measurement is merely 10 $k_{\rm{B}}~\mu$m$^{-2}$. Our estimates suggest that a single terahertz photon should be detectable at 300 mK using a 1 $\mu$m$^2$ size sample and a SQUID amplifier\cite{Fong:2012p2256, McKitterick:2013p2396, Muck:2003p3266}. We have also demonstrated the control of heat flow in graphene by both using the field effect and employing superconductors to suppress the thermal diffusion channel. These findings point the way to future experiments to probe both the fundamental and practical electronic thermal properties of this unique atomically thin material.

\section{Acknowledgments}
We acknowledge helpful conversations with P.~Kim, J.~Hone, E.~Henriksen, and D.~Nandi. This work was supported in part by (1) the FAME Center, one of six centers of STARnet, a Semiconductor Research Corporation program sponsored by MARCO and DARPA, (2) the US NSF (DMR-0804567), (3) the Institute for Quantum Information and Matter, an NSF Physics Frontiers Center with support of the Gordon and Betty Moore Foundation, and (4) the Department of Energy Office of Science Graduate Fellowship Program (DOE SCGF), made possible in part by the American Recovery and Reinvestment Act of 2009, administered by ORISE-ORAU under contract no. DE-AC05-06OR23100. We are grateful to G.~Rossman for the use of a Raman spectroscopy setup. Device fabrication was performed at the Kavli Nanoscience Institute (Caltech) and at the Micro Device Laboratory (NASA/JPL), and part of the research was carried out at the Jet Propulsion Laboratory, California Institute of Technology, under a contract with the National Aeronautics and Space Administration.


%

\clearpage

\begin{figure}[b!]
\begin{centering}

\includegraphics[width=2.0\linewidth]{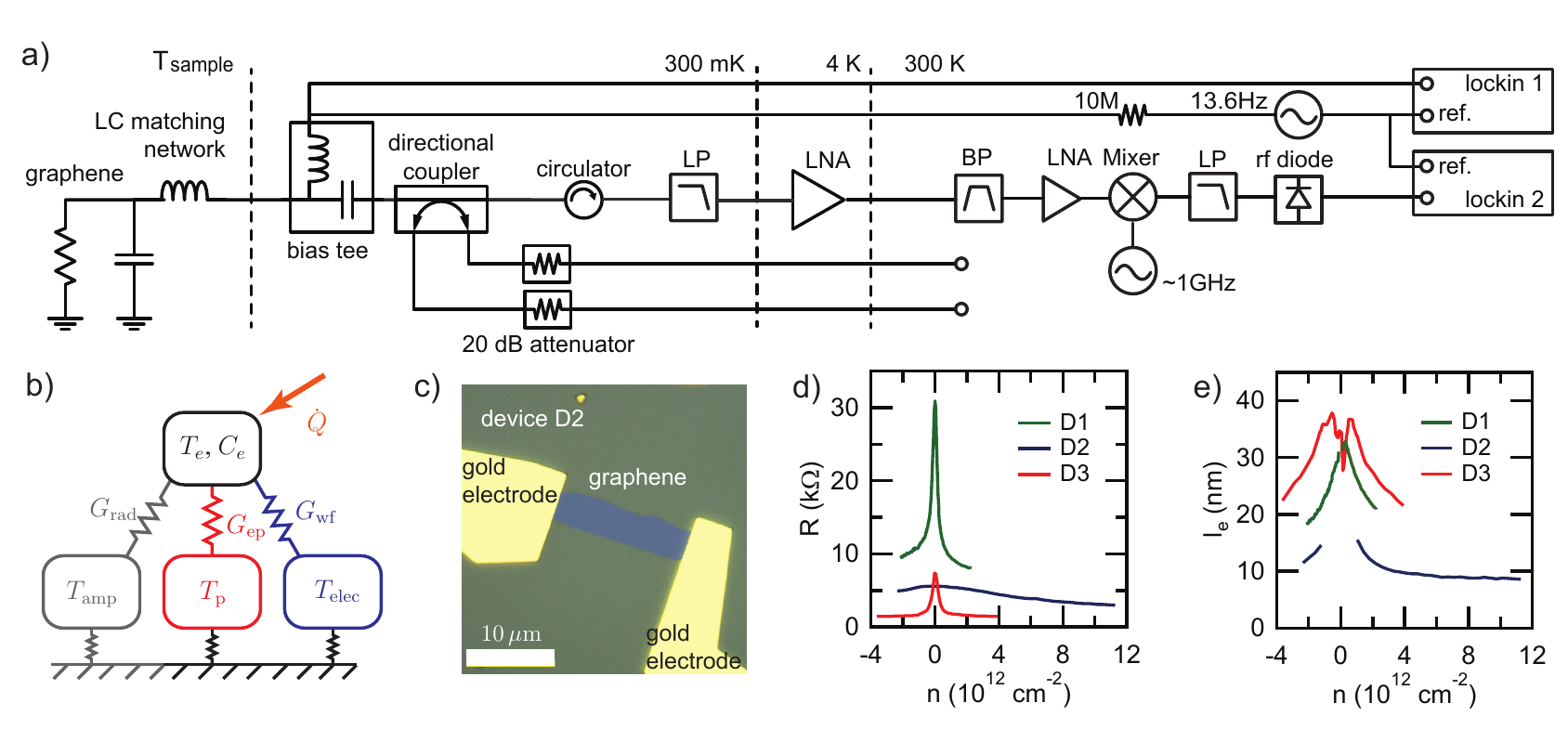} 
\end{centering}
\caption{ (a) Experimental setup for simultaneously measuring the thermal and electrical transport of a graphene device. (b) Graphene thermal model. Heat from the graphene electrons can flow out through two different channels:  electronic diffusion to the electrodes, $G_{\rm{wf}}$, and ep coupling, $G_{\rm{ep}}$. (c) Optical micrograph of device D2. (d) dc graphene resistances. (e) Electrical mean-free-path calculated using $l_{\rm{e}}=\sigma E_{\rm{F}}/ne^2v_{\rm{F}}$.}
\label{fig2}

\end{figure} 

\clearpage

\begin{figure}[b!]
\begin{centering}

\includegraphics[width=2.0\linewidth]{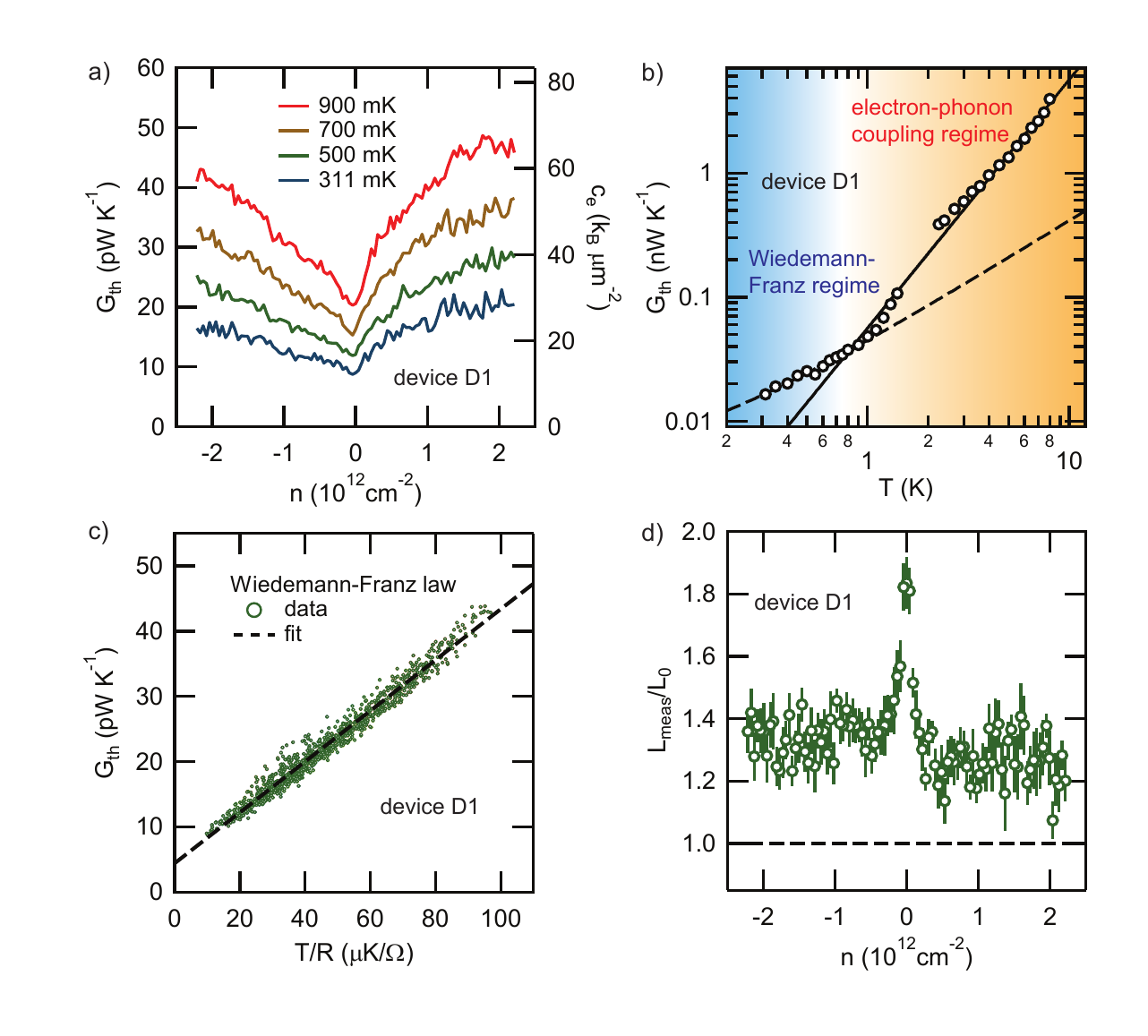} 
\end{centering}
\caption{ Data from device D1 with normal metallic electrodes: (a) The bipolar $G_{\rm{wf}}$ as a function of charge carrier density at various sample temperatures. (b) $G_{\rm{th}}$ data as a function of temperature at $n$ = -2.2$\times 10^{12}$ cm$^{-2}$. The solid line is the power law fit to the ep thermal conductance above 1.5 K while the dashed line is the best linear $T$ fit to the wf thermal conductance. The size of the data points represents the measurement error. (c) Wiedemann-Franz law in graphene. Each data point represents a measurement at a different temperature and charge carrier density. The fitted line is $G_{\rm{wf}} = \alpha\mathcal{L}_{\rm{meas}}T/R$ with a y-offset; $\mathcal{L}_{\rm{meas}} = 3.25\pm0.02\times 10^{-8}$ W$\Omega$K$^{-2}$ (d) The measured Lorenz number, $\mathcal{L}_{\rm{meas}}$, as a function of density from fitting of the Wiedemann-Franz law. For electrons with $n \leq -0.18\times 10^{12}$ cm$^{-2}$, averaged $\mathcal{L}_{\rm{meas}}/\mathcal{L} = 1.34\pm0.06$ while for holes with $n \geq +0.18 \times 10^{12}$ cm$^{-2}$, averaged $\mathcal{L}_{\rm{meas}}/\mathcal{L} = 1.26\pm0.07$.}

\end{figure} 

\clearpage

\begin{figure}[b]
\begin{centering}

\includegraphics[width=2.0\linewidth]{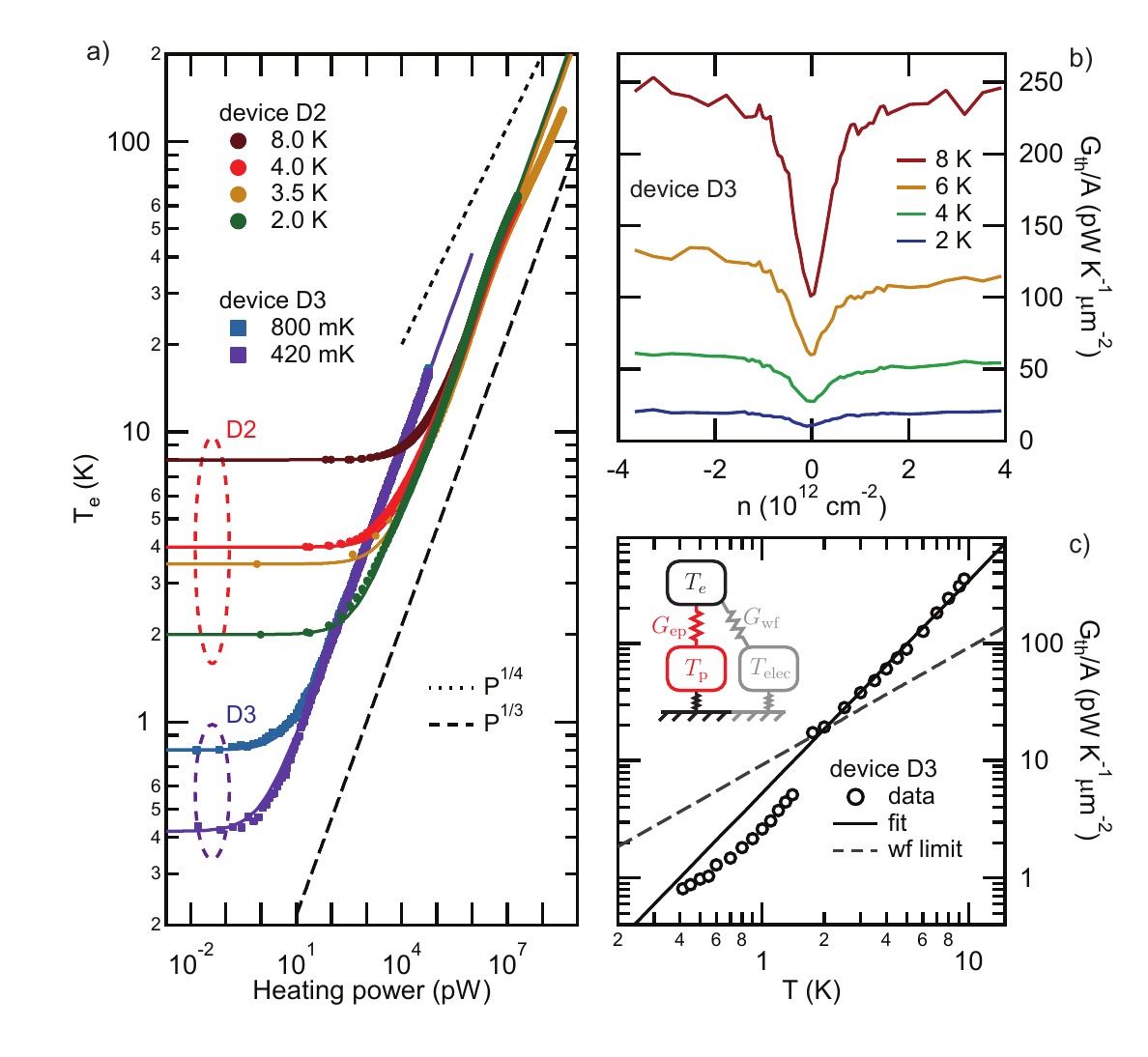} 
\end{centering}
\caption{ (a) The measured electron temperature versus dc heating power applied to the graphene devices at different phonon temperatures with $n = 28 \times 10^{12}$ cm$^{-2}$ for device D2 and $n = -2.2 \times 10^{12}$ cm$^{-2}$ for device D3. The solid lines are fits to Eqn.~\ref{eqn:HeatTransfer1}. The power law $T^3$ persists down to 420 mK for device D3 with NbTiN electrodes. (b) and (c) Data from device D3 with NbTiN electrodes: (b) the bipolar electron-phonon thermal conductance as a function of charge carrier density at various temperatures. (c) $G_{\rm{th}}$ data as a function of temperature at $n$ = -2.9$\times 10^{12}$ cm$^{-2}$. The solid line is the power law fit to the ep thermal conductance above 1.5 K while the dashed line is the calculated  $G_{\rm{wf}}=\alpha\mathcal{L}_0T/R$ using $R$ = 1770 $\Omega$. The offset between high and low temperature data is due to thermal cycling of the device on two cryostats.}

\end{figure}

\clearpage

\begin{figure}[b]
\begin{centering}

\includegraphics[width=2.0\linewidth]{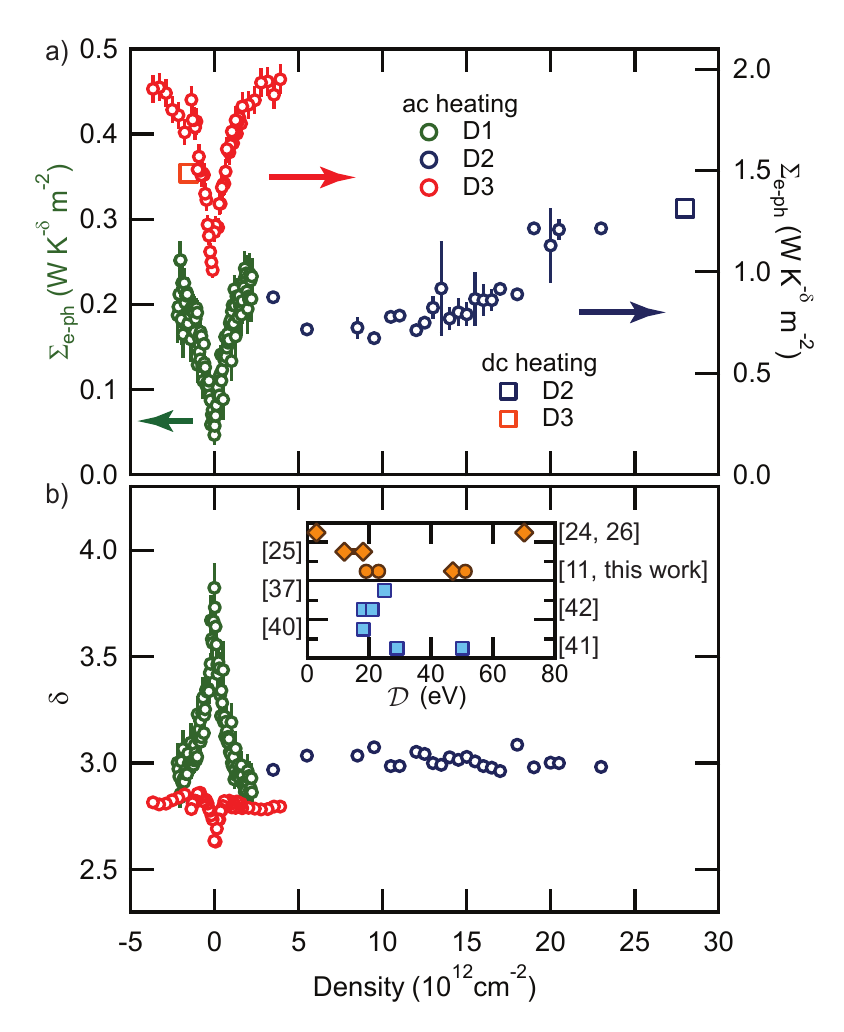} 
\end{centering}

\caption{ (a) Electron-phonon coupling parameter, $\Sigma_{\rm{ep}}$, and (b) the temperature power law, $\delta$, extracted from the thermal conductance data above 1.5 K as a function of charge carrier density in three measured devices. Inset: The values of the deformation potential, $\mathcal{D}$, in graphene from the literature: blue squares and orange diamonds represent data based on electrical and thermal-related measurements; orange circles represent the data from this report (see also Tab.~1).}

\end{figure}

\clearpage

\begin{figure}[b]
\includegraphics[width=2\linewidth]{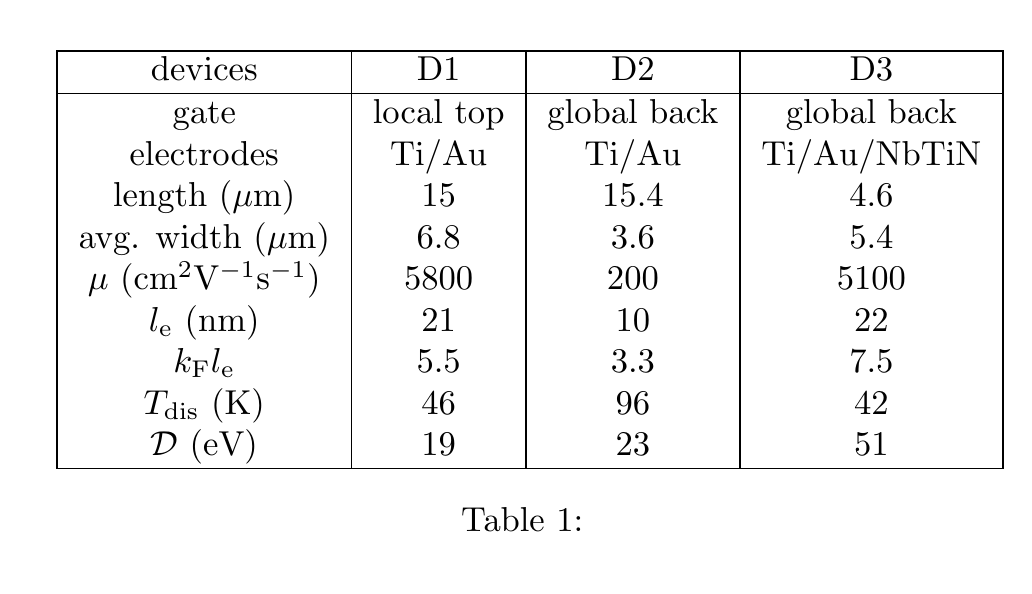} 

\caption{ Sample information and measured quantities. $\mu$ is the electronic mobility from fitting\cite{Dean:2010p2055}. $l_{\rm{e}}$ and $k_{\rm{F}}l_{\rm{e}}$ are quoted for data nearest to $n = 3.5\times 10^{12}$ cm$^{-2}$, corresponding to a Bloch-Gr\"{u}neisen temperature, $T_{\rm{BG}} = 2(s/v_{\rm{F}})(E_{\rm{F}}/k_{\rm{B}}) \simeq 101$ K. The disorder temperature is given by\cite{Chen:2012p2291} $T_{\rm{dis}} = hs/l_{\rm{e}}$. At this density, $T_{\rm{dis}} < T_{\rm{BG}}$ and $k_{\rm{F}}l_{\rm{e}} > 1$.}

\end{figure}

\end{document}